Calculations on aerostatic balloons made by the late Mr. Leonhard Euler, as they were found on his blackboard, after his death on 7 September 1783

E579 -- *Calculs sur les Ballons aérostatiques faits par le feu M. Léonard Euler, tels qu'on les a trouvés sur son ardoise, après sa mort arrivée le 7 septembre 1783*

Leonhard Euler

*Mémoires de l'académie royale des sciences de Paris,* Volume 1784, pp. 264-268.

Translated and Annotated[a]

by

Sylvio R. Bistafa[*]

May 2021

Foreword

This is an English translation of E579 in which the introductory remarks are in French, while Euler's original text is in Latin. By considering the balance of forces acting on a raising balloon on an isothermal atmosphere, namely the weight of the balloon, the buoyant force, and the aerodynamic drag force, Euler provides closed formulas for the calculation of the maximum altitude reached by the balloon, the altitude for which the velocity is maximum, the maximum velocity attained by the balloon, and the total ascending time.

______________________________

Admonishment

The experiment made in Annonay on June 5, 1783, by Mr. Montgolfier[b], showed the possibility of raising in the air bodies of a great capacity relative to their gravity, by filling them with an expandable fluid lighter than the air of the atmosphere, & whose elasticity is, nevertheless, in equilibrium with that of the air: scarcely was it known, when the *Savans* of Europe hastened to occupy themselves with an object which offered to almost all the Sciences new questions to be resolved, giving to anybody the hope of procuring new means of discoveries, & interested curiosity by the breath of real or chimerical applications, which, at first glance, presented the means of searching for an element which until then had been unknown to us.

Mr. Euler could not be informed until shortly before his death, of the discovery of Mr. Montgolfier. The idea of seeking the laws of the vertical motion of a globe rising in calm air, by virtue of the ascending force which it owes to its lightness, was the first which came to his mind: he immediately tried to apply calculations to this question; & when he was surprised by death, the blackboard on which he wrote with chalk, since he was almost deprived of his sight, was charged with these calculations, the last which have been made by this great man, also singular not only by the incredible number of his works, but also by the depth & the force of his genius.

---

[a] The Translator used the best of his abilities and knowledge to make this translation technically and grammatically as sound as possible. Nonetheless, interested readers are encouraged to submit suggestions for corrections as they see fitting.

[*] Corresponding address: sbistafa@usp.br

[b] The Montgolfier brothers are the inventors of the first practical balloon for flight. Their first demonstrated flight of a hot air balloon took place on June 4, 1783, in Annonay, France in front of an astonished crowd. Montgolfier brothers. (n.d). In Wikipedia. Retrieved May 19, 2021, from https://en.wikipedia.org/wiki/Montgolfier_brothers

The Academy of Sciences, to which the son of Mr. Euler, his successor in the place of Foreign Associate that he occupied among us, kindly sent a copy of these calculations, and hastened to publish them, like a precious monument which contains the last thoughts of one of the men who had done the most honor to the Sciences; as a singular proof of the vigor of the mind, which can still subsist a few hours before the moment, when an unknown cause will destroy the secret springs of intelligence & life; and finally, as an honor rendered to the author of the new discovery, since this same attempt at calculation proves the interest which it had stirred in one of those men whose suffrage is the most worthy recompense that genius can aspire to.

_______________________

Set the radius of an aerostatic globe $= a$, & weight $= M$, such that its volume $= \frac{4}{3}\pi a^3$, where $\pi$ is the periphery of the circle whose radius $= 1$. Set the height of a column of air $= k = 24000 \, ft$, approximately, & if we consider that the globe had reached the altitude $AM = x$, then, the air pressure[c] $= e^{-x/k}$.

Set the velocity of the globe at point $M = v$, & the weight of the globe of air $= N$, and, since the surface of the hemisphere $= \frac{\pi a^2}{2}$, then, the resistance at the point $M$ is equal to $\frac{v^2}{4g} \cdot \frac{\pi a^2}{2} \cdot \frac{N}{\frac{4}{3}\pi a^3} = \frac{3N}{8a} \cdot \frac{v^2}{4g}$, where $g$ is the height that a free body falls in one second[d].

The mechanical principles furnish this equation[e]: $2v\partial v = \frac{4g\partial x}{M} \cdot P$, in which $\partial x$ is an element of the altitude $Mm$ & $P$ is the soliciting force, that is composed with the expressions of the air, weight of the globe & resistance as[f]

---

[c] This is the so-called barometric law in which the atmospheric pressure and density decrease exponentially with altitude, under the simplifying assumption of an isothermal atmosphere.

[d] The height that a weight falls in on second is the acceleration of gravity divided by two; hence the actual acceleration of gravity is $2g$.

[e] Considering that $P$ is a force, then by the 2nd law, $P = \frac{M}{2g}\frac{dv}{dt}$, and the work done by this force is $Pdx = \frac{M}{2g}vdv$, which, indeed, can be written as $2vdv = \frac{4g}{M}Pdx$.

[f] In this expression, $Ne^{-x/k}$ corresponds to the buoyant force, which is given by the weight of the atmospheric air that is displaced by the globe at a height $x$, the second term is the weight of the balloon, and the third term is the aerodynamic drag force (the resistance). The resistance can be rewritten as $\frac{v^2}{2} \cdot \frac{\pi a^2}{2} \cdot \frac{1}{2g}\gamma_{air}e^{-x/k} = \frac{v^2}{2} \cdot \frac{\pi a^2}{2} \cdot \rho_{air}e^{-x/k}$, where $\gamma_{air}e^{-x/k}$ is the specific weight of the atmospheric air surrounding the balloon at a height $x$, $\rho_{air}$ is the density of the air in the balloon ($\rho_{air} = \frac{\gamma_{air}}{2g}$). Therefore, the resistance is seen to be given by the product of the dynamic pressure at height $x$, $\rho_{air}\frac{v^2}{2}e^{-x/k}$, and the projected area of the balloon in the horizontal plane $\frac{\pi a^2}{2}$. In fact, the resistance is also affected by the drag coefficient $C_D$, which is around 0.5 for spheres at a Reynolds number of about $10^4$.

$$P = Ne^{-x/k} - M - \frac{3N}{8a} \cdot \frac{v^2}{4g} e^{-x/k},$$

then,

$$2v\partial v = \frac{4g\partial x}{M}\left(Ne^{-x/k} - M - \frac{3N}{8a} \cdot \frac{v^2}{4g} e^{-x/k}\right),$$

or

$$2v\partial v = \frac{4g\partial x}{M}\left(4g\partial x \frac{N}{M} e^{-x/k} - 4g\partial x - \frac{3}{8a}\frac{N}{M} \cdot v^2 \partial x e^{-x/k}\right),$$

and by defining $\lambda = \frac{N}{M}$, then,

$$2v\partial v + \frac{3\lambda}{8a} \cdot v^2 e^{-x/k} \partial x = 4g\partial x(\lambda e^{-x/k} - 1),$$

which after integration and by setting $\frac{8a}{3\lambda} = b$, results in

$$v^2 e^{x/b} = \int 4g\partial x \left(\lambda - 1 - \frac{\lambda x}{k}\right) e^{x/b},$$

and can be rewritten as:

$$v^2 e^{x/b} = 4\lambda g \int e^{x/b} \partial x \left(\frac{\lambda - 1}{\lambda} - \frac{x}{k}\right) = \frac{4\lambda g}{k} \int e^{x/b} \partial x (f - x),$$

where $f = \frac{(\lambda - 1)k}{\lambda}$.

It is true that

$$\int e^{x/b} \partial x (f - x) = b(b + f)\left(e^{x/b} - 1\right) - be^{x/b} x,$$

then,

$$v^2 e^{x/b} = \frac{4\lambda gb}{k}\left[(b + f)\left(e^{x/b} - 1\right) - e^{x/b} x\right],$$

giving

$$v^2 = \frac{4\lambda gb}{k}\left[(b + f)\left(1 - e^{-x/b}\right) - x\right],$$

and this gives the velocity of the globe at any altitude.

For the determination of the maximum altitude that can be reached by the globe, let us consider that the velocity $v$ or its square vanishes at point $H$, and let us set $AH = h$, resulting in the following equation

$$(b + f)\left(1 - e^{-h/b}\right) - h = 0,$$

from which

$$b + f = \frac{h}{1 - e^{-h/b}} = \frac{he^{h/b}}{e^{h/b} - 1}.$$

Set $f = nb$, then, we have that $b + f = (n + 1)b$, and since $h$ is much less than $b$, we can consider, without much error, that $e^{h/b} - 1 = e^{h/b}$, and then, $b + f = b(n + 1) = h$, knowing that $b = \frac{8a}{3\lambda}$, &

$$n = \frac{3(\lambda - 1)k}{8a} \left[ since \ \left(\frac{\lambda - 1}{\lambda}\right)k = f = nb = \frac{8an}{3\lambda}\right].$$

For the ascending time, the equation $v \partial t = \partial x$ shows that

$$\partial x = \partial t \sqrt{\frac{4\lambda gb}{k}[h(1 - e^{-x/b}) - x]},$$

where $\partial t$ is the element of time. Hence, we have that

$$\partial t \sqrt{\frac{4\lambda gb}{k}} = \frac{\partial x}{\sqrt{[h(1 - e^{-x/b}) - x]}}.$$

Let us set $\partial t = \sqrt{\frac{k}{4\lambda gb}} \int \frac{\partial x}{\sqrt{[h - x]}}$, then, by integration, we have that

$$t = \sqrt{\frac{k}{4\lambda gb}}\left[C - 2\sqrt{(h - x)}\right] = \sqrt{\frac{k}{4\lambda gb}}\left[2\sqrt{h} - 2\sqrt{(h - x)}\right]$$

whence results in the ascending time throughout the distance $AM$ as $t = \sqrt{\frac{kh}{\lambda gb}}\left(1 - \sqrt{\frac{h-x}{h}}\right)$, and the total ascent time will be $\sqrt{\frac{kh}{\lambda gb}}$.

To determine the altitude for which the velocity is maximum, let us put $\partial[(b + f)(1 - e^{-x/b}) - x] = 0$, thence, $\frac{\partial x}{b}(b + f)e^{-x/b} = \partial x$, whence $\frac{b+f}{b} = e^{x/b}$, consequently, $x = b \cdot ln(b + f) - b \cdot ln(b) = b \cdot ln\ (n + 1)$, hence $AF = b \cdot ln\ (n + 1)$, and by substituting this result into the expression for the velocity, results in

$$v^2 = \frac{4\lambda gb^2}{k}\left[(n + 1)\left(1 - e^{-ln(n+1)}\right) - ln(n + 1)\right],$$

thence

$$v^2 = \frac{4\lambda gb^2}{k}[n - ln(n + 1)],$$

therefore, the maximum velocity at $F$ will be $= 2b\sqrt{\frac{\lambda g}{k}[n - ln(n + 1)]}$, or $2b\sqrt{\frac{\lambda ng}{k}}$, because the number $n$ is rather large.

Example: Set $a = 30\ ft$, $\lambda = \frac{N}{M} = 5$, $b = \frac{8a}{3\lambda} = 16$ & $n = \frac{f}{b} = \frac{3(\lambda-1)k}{8a} = \frac{3(5-1)24000}{8 \cdot 30} = 1200$, whence the maximum altitude $AH = h = b(n + 1) = 16 \cdot 1201 = 19216$, and the altitude for which the velocity is maximum $AF = b \cdot ln(n+1) = 16 \cdot ln(1201) = 113.5 \, ft$, the maximum velocity $2b\sqrt{\frac{\lambda n g}{k}} = 2 \cdot 16 \cdot \sqrt{\frac{5 \cdot 1200 \cdot 16.09}{24000}} = 64.18 \, ft/s$, with an ascending time $\sqrt{\frac{kh}{\lambda g b}} = \sqrt{\frac{24000 \cdot 19216}{5 \cdot 16.09 \cdot 16}} = 9'58''$.[g]

---

[g] The numerical results have been reviewed, and are slightly different from those that appear in the original publication.